\documentclass[twocolumn, amsmath, superscriptaddress, amsfonts,prb]{revtex4-2}
\usepackage{graphicx}
\usepackage{url}
\pdfoutput=1
\usepackage{mhchem}

\begin{document}

\title{Underscreening and related phenomena in concentrated electrolytes}

\author{Diana Shvydka}\email{diana.shvydka@osumc.edu}\affiliation{Department of Radiation Oncology, The Ohio State University Wexner Medical Center, Columbus, OH 43221, USA}
\author{Victor Karpov}\email{victor.karpov@utoledo.edu}\affiliation{Department of Physics and Astronomy, University of Toledo, Toledo,OH 43606, USA}

\begin{abstract}
We propose a heuristic model of underscreening phenomenon in high density Coulomb systems, such as concentrated electrolytes and electron hole conglomerates under ultra high dose rate (UHDR) radiation in biological tissues. It explains the data on screening length $L$ increasing with charge particle concentration and offers additional insights in understanding the conductivity and reduction potential of concentrated electrolytes. Also, it validates our current understanding of the FLASH radiation treatment of tumors (FLASH-RT) perceived as an analogous system. The underlying physics is that mutual binding creates diffusion barriers which suppress the concentration of mobile particles thus increasing the screening length. Also, they slow down the rates of chemical reactions responsible for generation of biologically active radicals which explains the sparing effect observed under UHDR.

\end{abstract}

%\blinddocument
%\begin{titlingpage}
\maketitle

%\end{titlingpage}
\section{Introduction}\label{sec:intro}

The structure of electrolytes constitutes a fundamental question important for multiple applications. The classic Debye-Hückel theory, valid for
dilute electrolytes, predicts that the interaction between two
charged flat surfaces in an electrolyte decays exponentially
with the surface separation. The corresponding decay length $\lambda$ , called
the Debye length, \cite{debyewiki} is given by
\begin{equation}\label{eq:deblen}\lambda (N)=\sqrt{\frac{\varepsilon k_BT}{4\pi Nq^2}}\end{equation}
where $\varepsilon$ is the dielectric constant of the medium (which may depend on ion concentration),
$k_B$ is the Boltzmann constant, $T$ is the temperature, $q$ and $N$ are respectively the ion charge and concentration. 

The corresponding interaction energy between two point charges distance $r$ from each other is described by the screened Coulomb potential,
\begin{equation}\label{eq:potential}\phi (r) =\frac{q^2}{r}\exp\left(-\frac{r}{\lambda}\right).\end{equation}

The applicability of classical results in Eqs. (\ref{eq:deblen}) and (\ref{eq:potential}) is limited to diluted electrolytes, i. e. 
\begin{equation}\label{eq:criterion}q^2N^{1/3}/\varepsilon \ll k_BT\end{equation}
as was extensively verified with many systems in physics of condensed matter, biophysics, and plasma physics. 

In recent decades, unexpectedly long decay lengths $\lambda (N)$ {\it increasing} with $N$ have been observed in concentrated
electrolytes [above the Debye-Hückel limits of Eq. (\ref{eq:criterion})] constituting the phenomenon known as “underscreening.” \cite{gebbie2015,smith2016,lee2017,gaddam2019,krucker2021,hartel2023,reinertsen2024,elliott2024,rondepierre2025,cross2026,tiger2026,baker2026}
A related observation is that $\lambda$ decreases in ionic solutions with increasing $T$. \cite{gebbie2015}

Among efforts to explain the underscreening, we mention charge correlations including ionic cluster formation, steric interactions and hydrodynamic effects, and slow out-of-equilibrium relaxation dynamics resembling that of glassy materials.
The above mentioned temperature dependence \cite{gebbie2015} was phenomenologically attributed to increase in the thermally driven effective concentration of detrapped
either free ions or correlated domains (quasiparticles), taking on the role of ions in traditional dilute electrolyte solutions. 

\section{A heuristic model of underscreening}\label{sec:model}
Here, we present a heuristic approach to underscreening over the entire range of dilute and concentrated electrolytes, simultaneously expanding to the classical electron-hole liquids in aqueous solutions under UHDR ionizing radiation. Our approach is based on accounting for the Coulomb interactions in a system of charged particles. Following the standard description of ionic solids, the Coulomb portion of binding energy per one particle is given by
\begin{equation}\label{eq:madelung} U= \frac{\mathfrak{M}q^2}{\varepsilon r}.\end{equation}
Here $r$ is the nearest neighbor distance and $\mathfrak{M}\sim 1-2$ is the Madelung constant. \cite{kittel1996,wikiMad,takenaka2024}.

The binding energy $U$ acts as a diffusion barrier. The probability to overcome it is described by the Boltzmann's exponent, 
\begin{equation}\label{eq:temp}p=\exp(-L)\quad {\rm with} \quad L\equiv\frac{U}{k_BT}=\frac{\mathfrak{M}q^2}{\varepsilon rk_BT}.\end{equation} 
The Coulomb system under consideration can behave as a low concentration plasma ($L\ll 1$), or ionic liquid ($L\gtrsim 1$), or ionic solid ($L\gg 1$). 

To estimate $L$ we approximate $r$ with the average interparticle distance under condition of
\begin{equation}\label{eq:n-r}N(4\pi r^3/3)=1.\end{equation} 
As a result, the concentration of `active' particles $N\exp(-L)$ capable of overcoming the Coulomb barriers becomes,
\begin{equation}\label{eq:mobpar}N_{\rm mobile}=N\exp\left[-\frac{\mathfrak{M}q^2}{\varepsilon k_BT}\left(\frac{4\pi N}{3}\right)^{1/3}\right]. \end{equation}
Note that $N_{\rm mobile}$ is a maximum at 
\begin{equation}\label{eq:maxcon}N=N_{\rm max}=\frac{81}{4\pi}\left(\frac{\varepsilon k_BT}{\mathfrak{M}q^2}\right)^3.\end{equation}

\begin{figure}[t!] 
\includegraphics[width=0.45\textwidth]{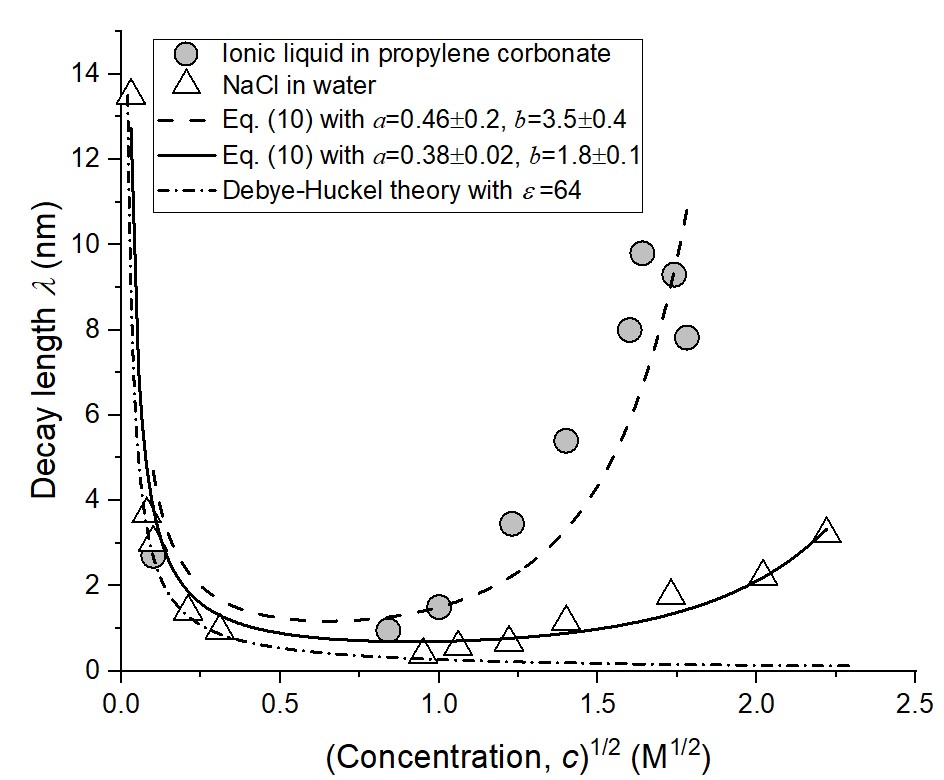} 
\caption{Fitting the data from Ref. \cite{smith2016} with Eq. (\ref{eq:fit}) and (\ref{eq:param}) Note that the concentration $c$ in Fig. \ref{fig:scrlen} is expressed in units mole/liter following teh original publication. }\label{fig:scrlen}
\end{figure}

It is our interpretation that the observed underscreening features can be described by Eq. (\ref{eq:deblen}) with $N_{\rm mobile}$ substituting for $N$,  
\begin{equation}\label{eq:scrlen}\lambda =\sqrt{\frac{\varepsilon k_BT}{4\pi N_{\rm mobile}q^2}}\end{equation} with $N_{mobile}$ given by Eq. (\ref{eq:mobpar}). 
reflecting our understanding that only mobile charges contribute to screening. We observe that the screening length of Eq.(\ref{eq:scrlen}) is a minimum at $N=N_{\rm max}$; hence the screening length is a minimum at that concentration. 

Our interpretation is verified by fitting the published data in Fig. \ref{fig:scrlen}  with Eq. (\ref{eq:scrlen}) written in the form 
\begin{equation}\label{eq:fit}\lambda =\frac{a}{\sqrt{N}}\exp(bN^{1/3})\end{equation}
containing two fitting parameters, 
\begin{equation}\label{eq:param} a=\sqrt{\frac{\varepsilon k_BT}{4\pi q^2}},\quad {\rm and}\quad b=\frac{\mathfrak{M}q^2}{\varepsilon k_BT}\left(\frac{4\pi}{3}\right)^{1/3}.\end{equation}  
The best fit values of $a$ and $b$ are given in Fig. \ref{fig:scrlen}. The corresponding ratios of these values are qualitatively consistent with the known dielectric permittivities $\varepsilon =64$ for ionic solution and $\varepsilon \approx 81$ for NaCl in water. Note that in our fitting procedure we kept $\mathfrak{M}=1$ to avoid any additional assumptions. Note that our determined best fit parameters $b$ correspond to $L\gtrsim 1$, i. e. ionic conglomerate forming Coulomb liquid.

A comment may be in order to better explain our proposed mechanism of underscreening that implies a spatially uniform distribution of screening charges characterized solely by their {\it average} concentration where no specific clusters are implied. The mechanism is that the Coulomb diffusion barriers {\it uniformly} increase throughout the system suppressing the average diffusivity of charges; hence, increase in screening length.

Overall, our work exhibits some synergy with other publications on underscreening: \\
(i) It emphasizes the role of Coulomb correlations. \\
(ii) It is conceptually close to the interpretation \cite{gebbie2015} attributing underscreening to decrease in concentration of mobile ions, which in our case is described by Eq. (\ref{eq:mobpar}). Using $N\sim 10^{21}$ cm$^{-3}$ from Fig. \ref{fig:scrlen}, the detrapping exponent in Eq. (\ref{eq:mobpar}) turns out to be of the order of 10, in agreement with the independent approach  where we have used $N\sim 10^{21}$ cm$^{-3}$ and $\varepsilon\sim 12$, and $\mathfrak{M}\sim 1$ from \cite{gebbie2015}. \\
(iii) Our interpretation can explain the observed slow relaxation screening lengths with characteristic times reminiscent of aging phenomena in glasses and other noncrystalline systems. \cite{cross2026} The explanation is that the barrier heights $E_a$ [quantifying the binding energies $U$ of Eq. (\ref{eq:madelung})] have random contributions due to disorder in the interparticle distances. Following similar observations in the physics of disordered systems,\cite{subedi2017,karpov2007,anderson1972} the related probabilistic distributions of barriers $E_a$ must be close to uniform. The corresponding distribution $\rho (t)\propto 1/t$ for random relaxation times $t\propto \exp (E_a/kT)$ then predicts slow relaxations in the material structure.

\section{Some related phenomena}\label{sec:relphen}Here, we apply our theory to some other properties of concentrated  electrolytes, such as e. g. the dc electric conductivity $\sigma$ or electrochemical cell potential (reduction potential). Another class of related phenomena can be found with biological tissues possessing high density of charge carriers generated under UHDR of ionizing radiation as explained next.\\

\subsection{Electric conductivity}\label{sec:ec} Among multiple attempts of describing the electrolyte conductivity, we would like to mention several recent publications \cite{gilliam2007,vila2005,zhang2020,bernard2023,vinogradova2025,budkov2026} presenting a variety of data and theoretical approaches. Trying to streamline our approach, we start with replacing the ionic concentration  $N$ in the equation for conductivity to that of of mobile ions, i. e. 
\begin{equation}\label{eq:cond}\sigma =N_{\rm  mobile} q\omega\end{equation} 
for a single componentdominated  electrolyte where $\omega$ is the ionic mobility. To account for possible effects of Coulomb correlations on mobility and following the published discussions \cite{vila2005} we assume
\begin{equation}\label{eq:mobil}\omega =\omega _0 -\frac{\omega _1}{\lambda}\end{equation}
where $\omega _0$ is the mobility unperturbed by the Coulomb correlations, and $\omega _1$ is a material constant describing the effect of such correlations over the screening length $\lambda$. (Here we neglect possible additional structure of $\omega _1$ described e. g. in Eqs. (5) and (9) of Ref. \onlinecite{vila2005}).

\begin{figure}[t!] 
\includegraphics[width=0.47\textwidth]{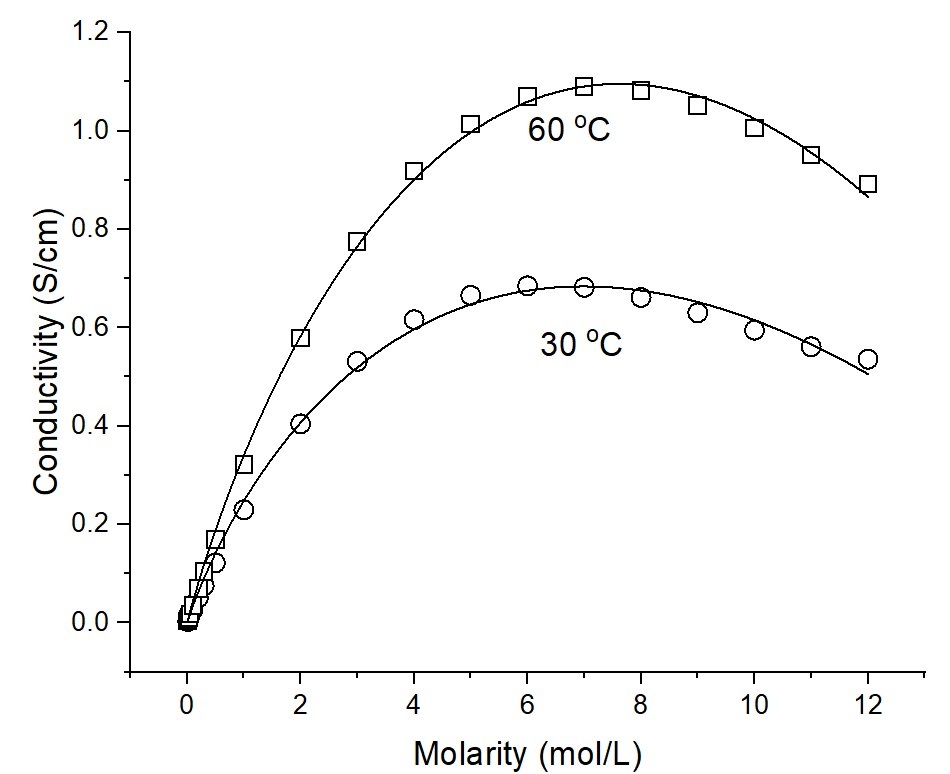} 
\caption{Fitting the data from Ref. \cite{gilliam2007} with the ansatz of Eq. (\ref{eq:condfinal}) wit the following sets of fitting parameters. For the 30 $^o$ C temperature data: $A=0.4\pm 0.03$, $B=0.16\pm 0.04$, $C=0.12\pm 0.01$.
For the 60 $^o$ C temperature data: $A=0.48\pm 0.02$, $B=0.49\pm 0.02$, $C=0.12\pm 0.008$.  }\label{fig:cond}
\end{figure}

Summarizing the above we arrive at the following expression for conductivity
\begin{equation}\label{eq:condfinal}\sigma = N\exp(-BN^{1/3})[A-CN^{1/2}\exp(-BN^{1/3}/2)],\end{equation}
which we will use for describing the data $\sigma (N)$ with three fitting parameters $A$, $B$, and $C$. [Note parenthetically that Eq. (\ref{eq:condfinal}) implies the molarity values in mol/Liter that corresponds to the particle concentration $\approx 600$ cm$^{-3}$.] We find the obtained fits in Fig, \ref{fig:cond} quite satisfactory and validating our description of Coulomb correlations in concentrated electrolytes. Given the best fit values of parameter $B$, this case yields $L\sim BN^{1/3}\sim 0.2-1$ falling rather into to plasma state domain.

%Following our approach, $N$ will represent the concentration of  , we will replace where the relaxation time $\tau$ remains vaguely defined. Simply looking at Eq. (\ref{eq:maxcon}) one can predict a maximum in conductivity at $N_{\rm max}$ given by Eq. (\ref{eq:maxcon}) as illustrated in Fig. \ref{fig:cond}. We observe that our best fits are not as perfect as those in Fig. \ref{fig:scrlen}; their maximum points do not even show up in the diagram. Such a discrepancy is expected because we are missing important information about the relaxation time $\tau$. Yet, our model provides the ratio of maximum conductivities at two different temperatures. Using Eq. (\ref{eq:maxcon}) one gets $(\sigma _1)_{\rm max}/(\sigma _2)_{\rm max}\approx (T_1/T_2)^3\approx 1.33$ in fair agreement with the data.\\

In fact, the conductivity data exhibit  significant variations in shape and values between different conditions and sources. We have checked that our fitting procedure provides satisfactory fits in most cases including shapes where the conductivity remains almost flat upon achieving it maximum or flattens in its part beyond the maximum point. We find it important to note that our approach is significantly based on the concept of Coulomb correlations that exhibit themselves through the exponentials in the final equation (\ref{eq:condfinal}). Enforcing those exponentials to disappear, say through setting $B=0$ in Eq. (\ref{eq:condfinal}), significantly deteriorates the fits quality.

\subsection{Reduction potential}\label{sec:rp}
Li ion batteries utilize energies generated by practical electrochemical systems beyond the applicability of dilute electrolyte theory.\cite{ko2023} However, the Nernst effect data relating the electrolyte concentration $N$ and the electrode electrostatic energy, 
\begin{equation}\label{eq:nernst}\Delta E_{Li,Li+}=(k_BT/n)\ln(N)\end{equation}  are considered informative where $n$ is the number of electrons involved in reduction. \cite{wikiNernst,orna1989}

\begin{figure}[h!] 
\includegraphics[width=0.37\textwidth]{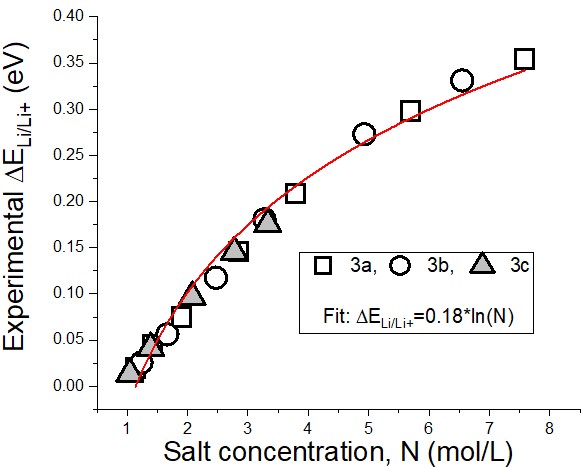} 
\caption{Fitting the data from Fig. 3a, 3b, and 3c of Ref. \cite{takenaka2024} with the fitting equation  $\Delta E_{Li,Li+}=7k_BT\ln(N)$ calculated in the diagram. }\label{fig:nernst}
\end{figure}

Representative data on Li related reduction are shown in Fig. \ref{fig:nernst} accommodating the information from three separate figures 3(a), 3(b), and 3(c) of the source publication [\onlinecite{takenaka2024}] that describe three electrolytes based on chemically different salts. Here, the three data sets are put on the same graph because they look similar and appealing towards a unified description. Indeed, a simple fit with just one parameter $n$ in Fig \ref{fig:nernst} describes these data well, although the measured Nernst potential remains to be several times stronger than expected. 

To understand the latter observation, we estimate the interionic coupling energy within a sphere of screening radius,
\begin{equation}\nonumber\int_0^\infty Nq^2\exp\left(-\frac{r}{\lambda}\right)4\pi rdr=k_BT\exp\left[\frac{\mathfrak{M}q^2}{\varepsilon k_BT}\left(\frac{4\pi N}{3}\right)^{1/3}\right]\end{equation}
where we have taken into account Eqs. (\ref{eq:deblen}) and (\ref{eq:mobpar}). That energy significantly exceeds $k_BT$. Using $\varepsilon \sim 10$, $N\sim 10^{21}$ cm$^{-3}$ and $\mathfrak{M}\sim 1$, one gets
\begin{equation}\label{eq:n}\frac{1}{n}=\exp\left[\frac{\mathfrak{M}q^2}{\varepsilon k_BT}\left(\frac{4\pi N}{3}\right)^{1/3}\right]\sim 10.\end{equation}
The latter figure is consistent with the value of fitting parameter $1/n =7$ in Fig. \ref{fig:nernst}; no other fitting parameters are needed here. Comparing Eqs. (\ref{eq:n}) and (\ref{eq:temp}) enables one to estimate $L=\ln (1/n)\approx 2$ referring the present case of reduction potential to the category of ionic liquids.

\subsection{Classical electron-hole liquid}\label{sec:ehl} The electrons and holes in water based biological tissues are solvated carrying along clouds of the coupled molecular structures. As a result, they possess much heavier effective masses behaving in many respects like ions. These quasi-ions are created in high concentrations under UHDR of ionizing radiation used for cancer treatments. A new FLASH modality of such treatments discovered in the past years exhibits rather unique properties where extremely short treatments maintain the conventional antitumor effectiveness while substantially reducing damage to normal tissues (sparing effect). \cite{vozenin2024}  

While the mechanism of FLASH modality remains to be sufficiently understood yet, it was conjectured \cite{shvydka2026,shvydka12026}  that the quasi-ions under UHDR can form the classical electron-hole liquid with relevant properties similar to those of concentrated electrolytes. In that system, suppression of diffusivities \cite{bernard2023} and chemical reactions \cite{borodin2020} slow down the generation rate of biologically active radicals leading to the sparing effect in normal tissues. (That mechanism does not apply to the structurally disordered cancer tissues where the quasi-ions concentration is limited due to their recombination.) In the framework of our heuristic theory here, the diffusivity suppression takes place when $N>N_{\rm max}$ in Eq. (\ref{eq:mobpar}); more elaborate analyses \cite{bernard2023,borodin2020} bring in concomitant interactions of steric and hydrodynamic nature. 

\section{conclusions}\label{sec:concl}
Our paper is based on a simplified mean-field description of the Coulomb interaction in an ensemble of charged particles presenting concentrated electrolyte. Our description proceeds from the typical solid state approach (Madelung constant, etc.) rather than some alternate pathways like Born equation  \cite{atkins2006} used in the physics of plasma. We have introduced a thermodynamic parameter $L$ [Eq. (\ref{eq:temp})] depending on the value of which our approach can describe either plasma ($L\ll 1$), or (Coulomb classical) liquid ($L\sim 1$), or ionic solid ($L\gg 1$). All three cases are presented in the above. To avoid any misunderstanding we would like to emphasize that our `$L$-classification' pertains to the ensemble of charged particles but not to the aggregate state of their underlying material: we can anticipate Coulomb liquid in regular liquid, plasma in ionic crystal, etc.   

Our description uniquely provides a simple analytical fitting for the underscreening effect data, electrolyte conductivity, and reduction potential. As the point of special significance for the authors of this manuscript, we would like to emphasize the here derived similarity between the properties of concentrated electrolytes and a hypothetical electron-hole liquid in biological tissues under UHDR. That similarity additionally validates our previously published conjecture of the classical electron-hole liquid as an important element underlying the mechanism of FLASH modality of radiation treatment of cancer.\cite{shvydka2026,shvydka12026}

\end{document}